# Discrepancy in the Upper Bound Mass of Neutron Stars

Sandeep Kumar S · Arun Kenath · C Sivaram

**Abstract** Observations have indicated that we do not see neutron stars (NS) of mass near the theoretical upper limit as predicted. Here we invoke the role of dark matter (DM) particles in star formation, and their role in lowering the mass of remnants eventually formed from these stars. Massive stars can capture DM particles more effectively than the lower mass stars, thus further softening the equation of state of neutron star. We also look at the capture of DM particles by the NS, which could further soften the upper mass limit of NS. The admixture of DM particles would be higher at earlier epochs (high z).

**Keywords** neutron stars · dark matter · star formation · early epoch

Sandeep Kumar S

Department of Physics and Electronics, Christ (Deemed to be University), Bengaluru, 560029, Karnataka, India

E-mail: sandeep141994@gmail.com

Arun Kenath

Department of Physics and Electronics, Christ (Deemed to be University), Bengaluru, 560029, Karnataka, India

Department of Physics, Christ Junior College, Bengaluru, 560029, Karnataka, India

E-mail: kenath.arun@cjc.christcollege.edu

C Sivaram

Indian Institute of Astrophysics, Bengaluru, 560034, Karnataka, India

E-mail: sivaram@iiap.res.in

# 1 Introduction

Prediction of the existence of stars supported by electron degeneracy pressure, led to startling discoveries in the field of astrophysics (Chandrasekhar, 1931). Landau anticipated the existence of stars supported by neutron degeneracy pressure in 1932 (Landau, 1932). Baade and Zwicky were the first to hypothesize that the remnant core left after a supernova explosion is a neutron star in 1934 (Baade and Zwicky, 1934). Discovered in 1967 by Jocelyn Bell Burnell and Antony Hewish as pulsars later identified as rapidly rotating neutron stars (Hewish et al., 1968). Neutron stars (NS) are astronomical objects with densities roughly of the order of $10^{17} - 10^{18}$ kg/m$^3$ (Hartle and Sabbadini, 1977; Goldman and Nussinov, 1989), roughly $10^8$ times higher than that of a white dwarf. Several models have been put forth to constrain the equation of state (EoS) of the interior of the neutron star. Since the prediction of neutron star's existence and discovery, more than a hundred EoS candidates have been suggested. But, only few have been realistic and successful in co-relating with the observations. One of the most massive pulsars measured is the PSR J0751+1807, with a mass of $2.1 M_\odot$ (Nice et al., 2005).

PSR J1748-2021B is the only neutron star to have mass of around $2.74 M_\odot$ with error bars. PSR B1957+10 and PSR J1311-3430 are the other objects to have a mass of $2.5 M_\odot$ (Lattimer, 2015). Apart from these objects, we do not observe neutron stars at the theoretically defined upper limit. The discrepancy between the observation and theory may be due to the presence of exotic matter particles such as hyperons, quarks etc. Strong gravity can be one of the reasons for lower mass limit observed at high densities (Capozziello et al., 2016). Determining a realistic upper bound mass limit for a neutron star is still an unsolved problem in the field of astronomy. This work invokes the presence of dark matter particles, mainly WIMPs (weakly interacting massive particles), inherently in the star-forming cloud at the time of formation of these stars, capture of dark matter particles by progenitor stars and also by neutron stars, and the effect of consequent capture of WIMPs on the maximum mass limit of neutron stars.

# 2 Neutron stars with inherent dark matter constituents

Newtonian gravity is effective in describing the white dwarf, but inadequate in the case of neutron stars. General relativity well describes the equation of state of neutron stars. The maximum limit for neutron star cores was first given by Tolman, Oppenheimer and Volkoff. No greater than $0.7 M_\odot$ neutron star core can exist in nature was their argument (Tolman, 1939; Oppenheimer and Volkoff, 1939). We find neutron stars much greater than $1 M_\odot$. Tolman, Oppenheimer and Volkoff did not consider the interaction of neutrons, thereby a softer mass limit was obtained. The first realistic EoS estimate was calculated on the work based on Rhoades and Ruffini (Rhoades and Ruffini, 1974). The better theoretical EoS estimate considering neutron-neutron scattering data is given by Kalogera and Baym (Kalogera and Baym, 1996). According to their estimates the upper limit mass is ~ $2.9 M_\odot$. The maximum mass limit of white dwarf has been worked out and is given as (Shapiro and Teukolsky, 1983):

$$M_{ch} \approx 0.78 \left(\frac{\hbar c}{G}\right)^{3/2} \left(\frac{1}{m_p}\right)^2 kg \qquad (1)$$

Where, $\hbar$ is the reduced Planck constant, $c$ is the speed of light, $G$ is the gravitational coupling constant and $m_p$ is the mass of the proton. The maximum upper limit mass for white dwarf is found out to be $1.44 M_\odot$ and is consistent with what is found in nature. Similarly, one might expect to find the maximum upper bound on the mass of a neutron star and observe it. But we do not observe neutron stars at the above limit. The maximum mass limit of neutron star has been worked out and is given as (Kippenhahn et al., 2012):

$$M_{NS} \approx 1.56 \left(\frac{\hbar c}{G}\right)^{3/2} \left(\frac{1}{m_n}\right)^2 kg \qquad (2)$$

Where $m_n$ is the mass of the neutron. The mass of a neutron is of the order of 1GeV. $M_{NS}$ varies as inversely proportional to the square of $m_n$.

A small admixture of dark matter particles at these cores, brings down value of the above mass limit. One of the possible explanations in the discrepancy could be the presence of these particles (Arun et al., 2018). The dark matter particle of interest is the WIMP. The energy of these particles is around $10 - 100$ GeV and they are considered stable (Arun et al., 2017). For the upper bound mass limit of $\approx 2.9 M_\odot$, a 1% of WIMP having mass of 10 GeV, present in the core lowers the limit to $\approx 2.4 M_\odot$. In simple words, if out of 100 particles (in NS), there is one DM particle of mass 10 GeV and 99 neutrons of mass 1 GeV, this implies that instead of 100 GeV bound mass, we now have effectively 109 GeV. So for 1% of DM present, the effective $m_n$ is increased by a factor of 1.09, and hence $M_{NS}$ is lower by a factor of $(1.09)^2 = 1.1881$. The change in the mass of NS's for varying WIMP masses is given in table (1).

Figure (1) shows the change in mass limit with DM fraction *f*. Dark matter particles could have been inherently present along with the baryons. These DM particles are considered as non-interacting with baryons and themselves, contributing only to gravitational pressure, effectively lowering the limit. This could be one of the plausible reasons in the mass discrepancy between the theoretical and observed neutron star masses. The neutron stars that have been observed lies in the lower half of the mass spectrum (Lattimer, 2012). This scenario is very likely for the stars formed in the early universe, since dark matter is said to constitute the framework for structure formation in the early universe (Arun et al., 2019), and the density goes as:

$$\rho = \rho_0 (1+z)^3 \qquad (3)$$

$\rho$ is the dark matter density at the epoch of formation of progenitor star, $\rho_0$ is the dark matter density at the present epoch and $z$ is the redshift.

| Mass of WIMPs | $M_{NS}$ |
|---|---|
| GeV | $M_\odot$ |
| 10 | 2.4051 |
| 20 | 1.9997 |
| 30 | 1.6887 |
| 40 | 1.4450 |
| 50 | 1.2505 |
| 60 | 1.0928 |
| 70 | 0.9631 |
| 80 | 0.8552 |
| 90 | 0.7645 |
| 100 | 0.6875 |

**Table 1** The change in the maximum mass limit of NS for a DM fraction of 1%

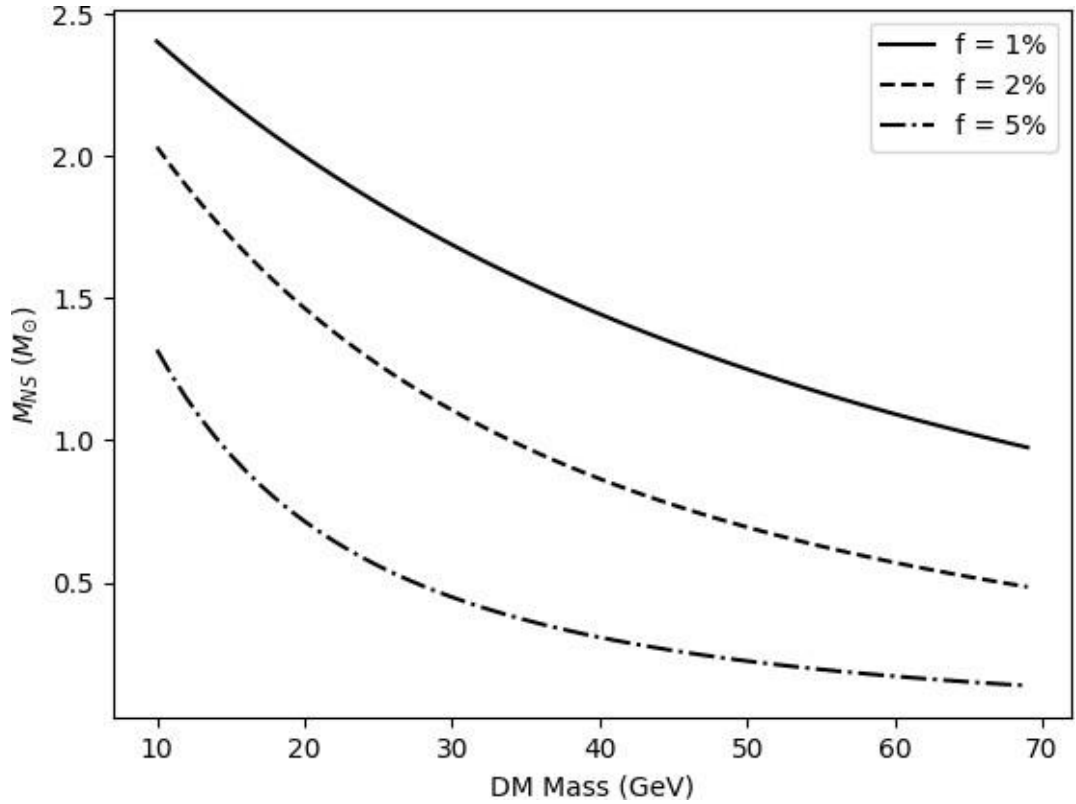

**Fig. 1** Change in $M_{NS}$ for different DM admixture.

# 3 Capture of dark matter particles by progenitor stars

Stars typically ranging from $8 - 25 M_\odot$ are massive stars, which undergo supernova explosions resulting in a neutron star or a black hole. A massive star of $25 M_\odot$ is considered here because we are interested in the upper bound mass limit of a NS. It can either end as a black hole or a neutron star, depending on the core mass left after the supernova explosion. Since these stars are massive they have higher chance of capturing WIMPs during the hydrogen and the helium burning stage. These stars roughly take around few million years to exhaust hydrogen (6.7 million years for a $25 M_\odot$ star) in the core and another hundred thousand years to exhaust helium (0.84 million years for a $25 M_\odot$ star) (Woosley et al., 2002). The other stages such as carbon, oxygen burning etc., in the cores have negligible timescales. The number of particles captured ignoring self-capture and self-annihilation effects is:

$$\frac{dN_\chi}{dt} = C_{(DM-Z)} \qquad (4)$$

Here $C_{(DM-Z)}$ is the rate of DM particle capture due to DM-nucleon interactions. The capture of WIMPs onto the star is given as (Press and Spergel, 1985; Kouvaris and Tinyakov, 2011):

$$C_{(DM-Z)} = \frac{8\pi^2 \rho_\chi}{3 m_\chi} \left(\frac{3}{2\pi v^2}\right)^{3/2} GMR v^2 \left(1 - e^{-3\epsilon_0/v^2}\right) g \qquad (5)$$

Equation (5) gives the number of particles captured per second. $\rho_\chi$ is the ambient dark matter density, $m_\chi$ is the mass of the dark matter particles, $M$ and $R$ are the mass and the radius of the star respectively, $v$ is the average dark matter dispersion velocity at that location. $\epsilon_0$ is the WIMP energy loss inside the star after a collision and is given as:

$$\epsilon_0 \approx 2 \frac{m_p GM}{m_\chi R} \qquad (6)$$

$g$ is the fraction of particles that undergoes scattering inside the star. For a sun like star, $g \approx 0.89\, \sigma_{\chi n}/\sigma_{crit}$. $\sigma_{\chi n}$ is the spin independent DM-nucleon cross section. $\sigma_{\chi n}$ for WIMP masses greater than 10 GeV is $10^{-51} \text{m}^2$ up to 35GeV from the recent XENON1T experiment (Aprile et al., 2017). The LUX dark matter experiment puts a constraint of $10^{-50} \text{m}^2$ at a mass of 33GeV (Carmona-Benitez et al., 2016). At a WIMP mass of 50GeV, $\sigma_{\chi n} \approx 10^{-50} \text{m}^2$ at 90% confidence limit (Akerib et al., 2017). From a detection point of view, $\sigma_{\chi n}$ is given as (Goodman and Witten, 1985):

$$\sigma_{\chi n} = \frac{m_\chi^2 M^2}{\pi (m_\chi + M)^2} |\mu|^2 \qquad (7)$$

Here $M$ is the mass of the target nucleus, $\mu$ is the weak scattering amplitude. It is clear from equation (7), $\sigma_{\chi n}$ proportional to $m_\chi^2$. Higher the mass of the DM particle, higher is the probability of interaction with the baryonic matter. The gamma ray excess measurements from Coma, Virgo, Fornax clusters indicates that the DM particle mass is in the range of 20 – 60GeV (Han et al., 2012). WIMP mass of 60GeV has an interaction cross section of $6.6 \times 10^{-48}$m$^2$ at 90% confidence limit (Figueroa-Feliciano, 2010). Therefore, a cross section of $10^{-48}$m$^2$ is considered. The condition required to trap the DM particle is $\sigma_{\chi n} < \sigma_{crit}$ and $\sigma_{crit}$ is given as (Kouvaris and Tinyakov, 2010):

$$\sigma_{crit} = \frac{m_p R^2}{M} \qquad (8)$$

For a star of mass $25 M_\odot$, the radius during the hydrogen burning stage is $9.17 R_\odot$. $\sigma_{crit}$ turns out to be $\sim 1.4 \times 10^{-39}$m$^2$. Therefore, from these values $g \approx 6.4 \times 10^{-10}$. From the above arguments $C_{(DM-Z)}$ is simplified to:

$$C_{(DM-Z)} = 3.7 \times 10^{30} \left(1 - e^{-3\epsilon_0/v^2}\right) \frac{\rho_\chi}{v \times m_\chi} \qquad (9)$$

Equation (9) gives the particles captured in a year. Consequently, the number of DM particles captured during the hydrogen burn phase ($t_{max}$), using equations (4) and (9) is:

$$N_\chi = 3.7 \times 10^{30} \left(1 - e^{-3\epsilon_0/v^2}\right) \frac{\rho_\chi \times t_{max}}{v \times m_\chi} \qquad (10)$$

The captured DM particles undergo collisions with other particles, lose energy and settle in the core forming a WIMP sphere of radius given as (Kouvaris and Tinyakov, 2011):

$$r_{th} = \left(\frac{9 k T_C}{8 \pi G \rho_C m_\chi}\right)^{1/2} \qquad (11)$$

$k$ is the Boltzmann constant, $T_C$ is the core temperature of the star, $\rho_C$ is the density at the core. For a $25 M_\odot$ star, $T_C = 3.81 \times 10^7$ K and $\rho_C = 3.81 \times 10^3$ kg/m$^3$ (Woosley et al., 2002). From these $r_{th}$ can be simplified to:

$$r_{th} = 2.72 \times 10^{-5} \left(\frac{1}{m_\chi}\right)^{1/2} meter \qquad (12)$$

For a DM mass of 60 GeV, the WIMP sphere radius is $\sim 8.3 \times 10^7$ m. The number of particles captured during hydrogen burning stage is $8.11 \times 10^{38}$ particles, for a WIMP mass of 60 GeV, $v = 270$ km/s and $\rho_\chi = 0.3 \times 10^6$ GeV/m$^3$ (local dark matter density) (Bovy and Tremaine, 2012). The number of DM particles captured is higher for stars in the galactic centre and globular clusters. Since the DM particles are confined to a small radius, even after the supernova explosion most of the captured dark matter particles are retained. These particles are squeezed to an even smaller radius at the onset of neutron star formation. The particle mass of DM is higher compared to neutrons, and they are considered to be non-interacting, contributing only to gravitational energy (if DM is a boson).

The maximum mass of a non-rotating and uniformly rotating neutron star is given as (Friedman and Ipser, 1987):

$$M_{nonrot} \approx 6.8 \left(\frac{\rho_0}{10^{17} kg m^{-3}}\right)^{-1/2} M_\odot \tag{13}$$

$$M_{rot} \approx 8.4 \left(\frac{\rho_0}{10^{17} kg m^{-3}}\right)^{-1/2} M_\odot \tag{14}$$

where $\rho_0$ is the mass density. From equations (13) and (14), the captured DM particles confined to a small radius, increases the mass density, in turn reducing the maximum mass of NS. There exists a limit beyond which the NS cannot be supported by degeneracy pressure against the gravitational collapse. This depends on the number and mass of the DM particles. If the DM particles are bosons, then the maximum number of bosons in the NS is given as (McDermott et al., 2012):

$$N_{chan}^{boson} \propto \left(\frac{1}{Gm_\chi^2}\right) \approx 1.5 \times 10^{34} \left(\frac{100 GeV}{m_\chi}\right)^2 \tag{15}$$

For a WIMP mass of 60 GeV, $N_{chan}^{boson} \approx 4.2 \times 10^{34}$. If the DM particles are fermions then, the maximum number of fermions in the NS is given as:

$$N_{chan}^{fermion} \propto \left(\frac{1}{Gm_\chi^2}\right)^{3/2} \approx 1.8 \times 10^{51} \left(\frac{100 GeV}{m_\chi}\right)^3 \tag{16}$$

For a WIMP mass of 60 GeV, $N_{chan}^{fermion} \approx 8.3 \times 10^{51}$. When $N_\chi > N_{chan}^{boson}$ or $N_{chan}^{fermion}$ the star collapses to a black hole. Number of DM particles captured by $25 M_\odot$ star during the hydrogen burning phase is ~ $10^{38}$ according to our consideration. If the DM is bosonic, then the star at the onset of neutron star formation collapses to a black hole $N_{self} > N_{chan}^{boson}$. In the early universe scenario, the progenitor stars could have inherent DM particles and further capture softens the upper bound mass of NS. This could be a possible explanation as to why we do not find neutron stars at the theoretically defined upper limit.

## 4 Capture of dark matter particles by neutron stars

We have already mentioned that the neutron stars are compact objects with high densities. Even though neutron stars have smaller surface area, the baryonic density and the gravitational force is immense, making them good accretors of matter. The DM particles captured by a neutron star for a certain time duration is given as (Zentner, 2009):

$$\frac{dN_\chi}{dt} = C_{(DM-Z)} + C_{(DM-DM)} N_\chi - C_{(A-DM)} N_\chi^2 \tag{17}$$

where $C_{(DM-Z)}$ is the rate of DM particle capture due to DM-nucleon interactions, $C_{(DM-DM)}N_\chi$ is the rate of DM particle capture due to DM self-interactions and $C_{(A-DM)}N_\chi^2$ denotes the particles lost due to annihilation of DM particles. Consider asymmetric DM, for which $C_{(A-DM)} = 0$. Then, equation (17) becomes as:

$$\frac{dN_\chi}{dt} = C_{(DM-Z)} + C_{(DM-DM)}N_\chi \qquad (18)$$

When solved for $N_\chi$, one obtains,

$$N_\chi = \frac{C_{(DM-Z)}}{C_{(DM-DM)}}(e^{C_{(DM-DM)} \times t} - 1) \qquad (19)$$

$C_{(DM-Z)}$ when general relativistic effects are considered, is given as (Kouvaris and Tinyakov, 2010):

$$C_{(DM-Z)} = \frac{8\pi^2 \rho_\chi}{3m_\chi}\left(\frac{3}{2\pi v^2}\right)^{3/2}\gamma v^2 (1 - e^{-3\epsilon_0/v^2})\xi g \qquad (20)$$

Equation (20) is the number of particles captured by the neutron star in a second. Here, $\gamma = GMR/\left(1 - \frac{2GM}{Rc^2}\right)$, $\xi$ is the Pauli blocking factor ~ 1, since $m_\chi > m_n$. $\epsilon_0$ is the WIMP energy loss inside the neutron star after a collision. In the case of a neutron star $\epsilon_0 \gg v^2/3$ (Güver et al., 2014). $g$ is the fraction of particles that undergo scattering inside the neutron star. $g = 1$, when $\sigma_{\chi n} > \sigma_{crit}$. For the condition $\sigma_{\chi n} < \sigma_{crit}$, $g = 0.45\, \sigma_{\chi n}/\sigma_{crit}$. But $\sigma_{crit} \approx m_n R^2/M$. For a neutron star of $2.9 M_\odot$ and radius of 10km, $\sigma_{crit} \approx 2.9 \times 10^{-50} m^2$, and $\sigma_{\chi n} \approx 10^{-48} m^2$. Since $\sigma_{\chi n} > \sigma_{crit}$, $g = 1$. Substituting these values in equation (20) gives:

$$C_{(DM-Z)} \approx \frac{2.3 \times 10^{35}}{v}\frac{\rho_\chi}{m_\chi} \qquad (21)$$

Equation (21) is the number of particles captured by the neutron star in a year. The presence of DM particles in the neutron star, aids in the capture of new dark matter particles and is given as:

$$C_{(DM-DM)} = \sqrt{\frac{3}{2}}\frac{\rho_\chi}{m_\chi}\sigma_{\chi\chi}\frac{v_{esp}(R)^2}{v}\langle\phi_\chi\rangle\beta\frac{erf(\eta)}{\eta} \qquad (22)$$

Equation (22) is the number of particles captured per second and also assumes uniform density of the neutron star, $\sigma_{\chi\chi}$ is the dark matter elastic scattering cross section and comes from the constraints by the observations of Bullet cluster and is $\sim 10^{-28} m^2$, $v_{esp}(R)$ is the escape velocity of the neutron star, that is $\sqrt{(2GM/R) - (GM/Rc)^2}$. For our consideration, $v_{esp}(R) \sim 0.8c$. $v$ is the average velocity of the dark matter particles at that location. $\phi_\chi \sim 1$ and $\frac{erf(\eta)}{\eta} \sim 1$, $\eta \equiv \sqrt{3/2}\, v_N/v$. $v_N$ is the velocity of the neutron star in the galaxy. Therefore, equation (22) becomes,

$$C_{(DM-DM)} \approx \frac{1.5 \times 10^{-3}}{v} \frac{\rho_\chi}{m_\chi} \tag{23}$$

Equation (23) gives the number of particles captured per year. On comparing equation (21) and (23), $C_{(DM-DM)}$ is several orders magnitude smaller than $C_{(DM-Z)}$. The solution for equation (18) now becomes:

$$N_\chi = C_{(DM-Z)} t \tag{24}$$

Substituting equation (21) in equation (24), we obtain:

$$N_\chi \approx (2.3 \times 10^{35}) \frac{\rho_\chi \times t}{v \times m_\chi} \tag{25}$$

Number of particles captured for a WIMP mass of 60 GeV, $v = 270$ km/s, $t = 10^9$ years (billion years) and $\rho_\chi = 0.3 \times 10^6$ GeV/m³ is ~ $4.3 \times 10^{42}$. The number of particles captured depends on the mass, density, interaction cross section and velocity of WIMPs. If sufficient number of particles are captured, the WIMP sphere overcomes the gravitational pull of the NS and self-gravitates. For this condition to be achieved, the number of WIMPs in the thermal radius must exceed that of the baryonic matter in the same volume and is given as:

$$N_{self} \approx 4.8 \times 10^{41} \left(\frac{100 GeV}{m_\chi}\right)^{5/2} \left(\frac{T}{10^5 K}\right)^{3/2} \tag{26}$$

For a WIMP mass of 60 GeV and $T = 10^5$ K, $N_{self} \approx 1.72 \times 10^{42}$ particles. But the Chandrasekhar limit for bosons of mass 60 GeV from equation (15) is $4.2 \times 10^{34}$. Since, $N_{self} > N_{chan}^{boson}$, the WIMP sphere collapses even before it enters the self-gravitational stage. From our above estimate, the number of WIMPs captured by a NS of $2.9 M_\odot$ in its lifetime is ~ $10^{42}$, which is much greater than the Chandrasekhar limit. In ten years since the formation of the NS, it would have captured enough DM to surpass the Chandrasekhar limit and begins to collapse to a black hole. The collapse time for a neutron star to become a black hole is given as (Bramante et al., 2013):

$$t_{NScollapse} \approx \frac{v_s^3}{4\pi \rho_b G^2 M_{BHinitial}} \tag{27}$$

$v_s$ is the speed of sound of the NS and is considered to be $c/\sqrt{3} \approx 0.6c$. $\rho_b$ is the baryonic density of the neutron star and calculated to be $1.4 \times 10^{18}$ kg/m³ for above considered mass and radius. $M_{BHinitial}$ is the initial mass of the black hole.

$$M_{BHinitial} = m_\chi N_{chan}^{boson} \tag{28}$$

For 60 GeV WIMP mass, $M_{BHinitial} = 4.48 \times 10^9$ kg (Sivaram et al., 2018). The collapse time is ~ $5.3 \times 10^8$ years. It roughly takes few hundred million years for the NS to collapse to black hole. The BH formed starts to evaporate via Hawking radiation and this time scale is given by:

$$t_{Hawking} \propto G^3 M_{BHinitial}^3 / \hbar c^4 \qquad (29)$$

The evaporation time is ~ 15 years. The evaporation of BH destroys the NS leaving nothing behind. If WIMPs are fermions, then higher number of particles are required for the collapse. This is due to fermi momentum which opposes the gravitational collapse.

## 5 Conclusions

There is overwhelming evidence for the existence of dark matter. The dark matter is said to have been prominently involved in the structure formation during the early universe, and a high probability that these dark matter particles could have been inherently present in the progenitor stars and hence, affecting the degenerate core mass. Also, massive stars capture dark matter particles more effectively than the lower mass stars, this could further soften the EoS of NS. From the detection point of view, to observe the NS is difficult at the said theoretical upper limit, because of the formation of BH in a few hundred million years (if DM is bosonic). The BH formed evaporates quickly and destroys the NS. The capture of particles is very prominent where the DM density is statistically higher than the ambient DM density, especially in the galactic centers and the globular clusters. From the above considerations and calculations, the presence of these DM particles along with other exotic particles in strong gravity regime is plausible, especially at the earlier epochs.

**Acknowledgments**

We would like to thank Blesson Mathew for valuable insights and constant encourage- ment. We would also take this opportunity to thank Ujjwal, Gayathri, Renée, Veena, Meghana and Krishna for their support during this work.